\documentstyle[12pt]{article}
\newcommand{\be}{\begin{equation}}
\newcommand{\ee}{\end{equation}}
\date{}
\begin{document}
\baselineskip 24.1pt plus 0.2pt minus 0.1pt
\title{Neutrino masses and mixing parameters in a left-right model with mirror fermions }

\author{\small R. Gait\'an$^2$, A. Hern\'andez-Galeana$^1$\thanks{With
support from EDI, COFAA and SNI}, J. M.
Rivera-Rebolledo$^1$\thanks{With support from COFAA-IPN, EDD and
SNI}\\
\small and P. Fern\'andez de C\'ordoba$^3$\\
\small  "Interdisciplinary Modeling Group, InterTech."                                 \\
\small  1. Departamento de F\'{\i}sica,\\
\small Escuela Superior de F\'{\i}sica y Matem\'{a}tica, I.P.N., \\
\small U.P. Adolfo L. Mateos, M\'{e}xico D.F., 07738, M\'{e}xico \\
\small                                                       \\
\small 2. Centro de Investigaciones Te\'oricas, FES, UNAM,\\
\small Apartado Postal 142, Cuatitl\'an-Izcalli, Estado de M\'exico,\\
\small C\'odigo postal 54700, M\'exico.\\
\small                                   \\
\small 3. Departamento de Matem\'atica Aplicada, \small
Universidad Polit\'ecnica de Valencia,\\
\small                           Valencia, Spain.}

\maketitle

\begin{abstract}
In this work we consider a left-right model containing mirror
fermions with gauge group SU(3)$_{C} \otimes $SU(2)$_{L} \otimes
$SU(2)$_{R} \otimes $U(1)$_{Y^\prime}$. The model has several free
parameters which here we have calculated by using the recent
values for the squared-neutrino mass differences. Lower bound for
the mirror vacuum expectation value helped us to obtain crude
estimations for some of these parameters. Also we estimate the
order of magnitude of the masses of the standard and mirror
neutrinos.

\end{abstract}


\section{Introduction}

The understanding of the properties of neutrinos is a problem that
has required an immense experimental and theoretical effort. There
is now convincing evidence that neutrinos have non-standard
properties: they have masses and their flavor states mix, leading
to processes like neutrino oscillations.

There are various features that make neutrinos specially
interesting. The smallness of neutrino masses is probably related
to the fact that neutrinos are completely neutral, i.e., they
carry no electric charge which is exactly conserved, and are
Majorana particles with masses inversely proportional to the large
scale where lepton number (L) conservation is violated.
\cite{yanagida} The relation with L non-conservation and the fact
that the observed neutrino oscillation frequencies are compatible
with a large scale for L non-conservation, establish connection
with Grand Unified Theories (GUT's). Then, neutrino masses and
mixing offer a different perspective on the problem of flavor and
the origin of fermion masses.

Another modern interest in neutrinos arises from their
significance in astrophysics and cosmology \cite{fukugita} and the
possible non-negligible contribution of neutrinos to hot dark
matter in the Universe.

An alternative explanation of small neutrino masses comes from the
concept of extra dimensions beyond the three ones that we know
\cite{aranda}. It has been suggested that right-handed neutrinos
(but not other particles of the Standard Model) experience one or
more of these extra dimensions. The right-handed neutrinos then
only spend part of their time in our world, leading to apparently
small neutrino masses.

From the experimental study of atmospheric and solar neutrinos we
have considerably improved our knowledge. The idea of neutrino
oscillations have gained support from the Japanese experiment
Super-Kamiokande \cite{superkamiokande} which in 1998 showed that
there was a deficit of muon neutrinos reaching Earth when cosmic
rays strike the upper atmosphere. These results were interpreted
as muon neutrinos oscillating into tau neutrinos, but they were
not detected.

Recently The Sudbury Neutrino Observatory (SNO)\cite{sno} in
Canada had confirmed that the solar neutrino deficit is due to
neutrino oscillations and not to a flaw in the model of the Sun:
the total neutrino flux is in agreement with the solar model but
only about one third arrives on Earth as $\nu_{e}$, while the
remaining part consists of other kinds of neutrinos, presumably
$\nu_{\mu}$ and $\nu_{\tau}$.

The KamLAND experiment has established that $\bar{\nu_e}$ from
reactors show oscillations over an average distance of about 180
Km which are perfectly compatible with the frequency and mixing
angle corresponding to one of the solutions of the solar neutrino
problem (the Large Angle solution) \cite{kamland}.

In this paper we use a model with gauge group SU(3)$_{C} \otimes
$SU(2)$_{L} \otimes $SU(2)$_{R}\\ \otimes $U(1)$_{Y^\prime}$ which
contains mirror fermions. In Section 1 we give an introduction to
the problem that we deal with; in Sec. 2 we present the model and
discuss the symmetry breaking part with the help of two scalar
doublets. Sec. 3 is devoted to write the Yukawa couplings for
Majorana and Dirac neutrinos, although we consider only these last
ones in calculations; the neutrino mass matrix is shown in terms
of three parameters; we diagonalise it and get expressions for the
physical neutrino masses. The use of their experimental values and
bounds gives us approximate numbers for the involved parameters.

\section{The model}
The gauge group symmetry is \cite{leftright} $G$ = SU(3)$_{C}
\otimes $SU(2)$_{L} \otimes $SU(2)$_{R} \otimes $U(1)$_{Y^\prime}$
and the content of fermions assumes the ordinary quarks and
leptons including the mirror counterpart, where the leptonic
sector is assigned under the group $G$ in the form:







$ l_{iL} \sim (1, 2, 1, -1)_{L}$,

$\nu_{iR} \sim (1, 1, 1, 0)_{R}$,

$e_{iR} \sim (1, 1, 1, -2)_{R}$,

$\hat{\nu_{iL}} \sim (1, 1, 1, 0)_{L}$,

$\hat{e_{iL}} \sim (1, 1, 1, -2)_{L}$,

$\hat{l_{iR}} \sim (1, 1, 2, -1)_{R}$,

where $i$ = 1,2,3 is the family index, the numbers in parenthesis
are the quantum numbers of the fermionic fields under the groups
SU(3)$_C$, SU(2)$_L$, SU(2)$_R$, and U(1)$_{Y^\prime}$,
respectively; the last entry corresponds to the hypercharge
($Y^{\prime}$), with the electric charge defined by $Q$ = T$_{3L}$
+ T$_{3R}$ + $\frac{Y^{\prime}}{2}$.

The "Spontaneous Symmetry Breaking" (SSB) is proposed to be
achieved in the stages
 \be \label{ecua} \mbox{G}\longrightarrow
\mbox{G}_{SM} \longrightarrow \mbox{SU(3)}_C \otimes
\mbox{U(1)}_Q\, \ee where G$_{SM}$ = SU(3)$_C \otimes $SU(2)$_L
\otimes $U(1)$_Y$ is the "Standard Model" group symmetry, and
$\frac{Y}{2}$ = $T_{3R}$ + $\frac{Y^\prime}{2}$.

The Higgs sector used to induce the SSB in eq.(\ref{ecua})
involves the scalar fields

\be
\Phi = (1, 2, 1, 1)
\ee

\be \hat{\Phi} = (1, 1, 2, 1)\ \ee where the entries correspond to
the transformation properties under the symmetries of the group
$G$, with the "Vacuum Expectation Values" (VEV's)

\begin{equation}
<\Phi> = \frac{1}{\sqrt{2}}
           \left(
            \begin{array}{cr}
             0\\
             v
             \end{array}
             \right),
\end{equation}

\begin{equation}
<\hat{\Phi}> = \frac{1}{\sqrt{2}}
               \left(
               \begin{array}{cr}
               0\\
               \hat{v}
               \end{array}
               \right)
\end{equation}
Here the VEV's $v$ and $\hat{v}$ are related to the masses of the
charged gauge bosons $W$ and $\hat{W}$ through $M_{W}$ =
$\frac{1}{2} g_L v$; $M_{\hat{W}}$ = $\frac{1}{2} g_R \hat{v}$,
with $g_L$ and $g_R$ being the coupling constants of SU(2)$_L$ and
SU(2)$_R$, and $g_L$ = $g_R$ if we demand $L$-$R$ symmetry.

\section{Neutrino mass matrix and mixing para- \\meters}
At this point we describe how to obtain the mass terms for the
sector of neutrinos (ordinary and mirror).

With the fields of fermions introduced in the model, we may write
the gauge invariant Yukawa couplings:

\be h_{ij} \overline{\hat{\nu}_{iL}} \nu_{jR} + \lambda_{ij}
\overline{l}_{iL} \tilde{\Phi} \nu_{jR} + \eta_{ji} \hat{l}_{jR}
\tilde{\hat{\Phi}} \hat{\nu}_{iL} + h.c., \ee
with
$\tilde{\Phi}$
= i $\sigma_2 \Phi^{*}$ and $\tilde{\hat{\Phi}}$ = i $\sigma_2
\hat{\Phi}^{*}$, where $h_{ij}$, $\lambda_{ij}$ and $\eta_{ij}$
are Yukawa coupling constants.

When $\Phi$ and $\hat{\Phi}$ acquire VEV's we get the Dirac type
neutrino mass terms

\be
h_{ij} \overline{\hat{\nu}_{iL}} \nu_{jR} + \frac{v}{\sqrt{2}}
\lambda_{ij} \overline{\nu_{iL}} \nu_{jR} +
\frac{\hat{v}}{\sqrt{2}} \eta_{ij} \overline{\hat{\nu}_{iL}}
\hat{\nu}_{jR} + h.c.
\ee

which may be written in the form:
\be
\overline{\Psi^{0}_{\nu L}}
M^{0}_{\nu}\Psi^{0}_{\nu R} + h.c.
\ee
with
\be M^{0}_{\nu} =
\left(
\begin{array}{cc}
\bf{\frac{v}{\sqrt{2}} \lambda}&\bf{0}\\
\bf{h}&\bf{\frac{\hat{v}}{\sqrt{2}} \eta}
\end{array}
\right)
\ee

\be \Psi^{0T}_{\nu L} = (\nu_{1 L}, \nu_{2 L}, \nu_{3 L},
\hat{\nu}_{1 L}, \hat{\nu}_{2 L}, \hat{\nu}_{3 L})
\ee

\be
\Psi^{0T}_{\nu R} = (\nu_{1 R}, \nu_{2 R}, \nu_{3 R},
\hat{\nu}_{1 R}, \hat{\nu}_{2 R}, \hat{\nu}_{3 R})
\ee

and $\bf{h}$, $\bf{\lambda}$ and $\bf{\eta}$ are matrices of
dimension $3 \times 3$.

The standard and mirror neutrino mass matrix $M_D$ is related to
the Dirac mass matrix $M^{0}_{\nu}$ by a bi-unitary transformation
through the equation:
\be
\label{mass1}
U^{+}_{L}M^{0}_{\nu} U_{R}
= M_D
\ee
where

\be
\label{mass2}
M_D = \left(
\begin{array}{cc}
m &0\\
0 &\hat{m}
\end{array}
\right),
\ee

and

\be
\label{rive}
U_{L,R} = \left(
\begin{array}{cc}
A &B\\
C &D
\end{array}
\right)_{L,R} = \left(
\begin{array}{cccccc}
a_{11} &a_{12} &a_{13} &b_{11} &b_{12} &b_{13}\\
a_{21} &a_{22} &a_{23} &b_{21} &b_{22} &b_{23}\\
a_{31} &a_{32} &a_{33} &b_{31} &b_{32} &b_{33}\\
c_{11} &c_{12} &c_{13} &d_{11} &d_{12} &d_{13}\\
c_{21} &c_{22} &c_{23} &d_{21} &d_{22} &d_{23}\\
c_{31} &c_{32} &c_{33} &d_{31} &d_{32} &d_{33}
\end{array}
\right)_{L,R} \ee

In matrix form we write : \be \left( \bar{\nu}_{L},
\bar{\hat{\nu}}_{L} \right) \left(
\begin{array}{cc}
\frac{v}{\sqrt{2}} \lambda &0\\
h       &\frac{\hat{v}}{\sqrt{2}} \eta
\end{array}
\right)
\left(
\begin{array}{c}
\nu_{R}\\
\hat{\nu}_{R}
\end{array}
\right)
\ee

Also, in eq.(\ref{rive}), $a_{12}$ is the mixing angle between
$\nu_e$ and $\nu_{\mu}$, that is, $a_{12}$ = $\theta_{12}$, etc.

As a first approximation we assume the mass matrix in the form \be
M^{0}_{\nu} = \left(
\begin{array}{cccccc}
0 &\frac{\lambda v}{\sqrt{2}} &\frac{\lambda v}{\sqrt{2}} &0 &0 &0\\
\frac{\lambda v}{\sqrt{2}} &\frac{\lambda v}{\sqrt{2}} &-\frac{\lambda v}{\sqrt{2}} &0 &0 &0\\
\frac{\lambda v}{\sqrt{2}} &-\frac{\lambda v}{\sqrt{2}} &\frac{\lambda v}{\sqrt{2}} &0 &0 &0\\
h &h &h &0 &\frac{\eta \hat{v}}{\sqrt{2}} &\frac{\eta \hat{v}}{\sqrt{2}}\\
h &h &h &\frac{\eta \hat{v}}{\sqrt{2}} &0 &\frac{\eta \hat{v}}{\sqrt{2}}\\
h &h &h &\frac{\eta \hat{v}}{\sqrt{2}} &\frac{\eta
\hat{v}}{\sqrt{2}}&0
\end{array}
\right)
\ee

The simple way to find the physical squared neutrino masses is by
diagonalization of the mass matrix $M^{0}_{\nu} M^{0 +}_{\nu}$ (or
$M^{0 +}_{\nu} M^{0}_{\nu}$) using eq.(\ref{mass1}). This gives
the following eigenvalues:

\be
m_1^2 = \lambda^2 v^2,
\ee

\be
m_2^2 = 2 \lambda^2 v^2,
\ee

\be
m_3^2 = \frac{1}{2} \left( 9 h^2 + \lambda^2 v^2 + 2 \eta^2
\hat{v}^2 - \sqrt{-16 \lambda^2 \eta^2 v^2 \hat{v}^2 + \left( 9
h^2 + \lambda^2 v^2 + 2 \eta^2 \hat{v}^2 \right)^2} \right),
\ee

\be
m_4^2 = m_5^2 = \frac{\eta^2 \hat{v}^2}{2},
\ee

\be
m_6^2 = \frac{1}{2} \left( 9 h^2 + \lambda^2 v^2 + 2 \eta^2
\hat{v}^2 + \sqrt{-16 \lambda^2 \eta^2 v^2 \hat{v}^2 + \left( 9
h^2 + \lambda^2 v^2 + 2 \eta^2 \hat{v}^2 \right)^2} \right).
\ee

From here one gets,

\be \label{matrix} m_{2}^{2} - m_{1}^{2} \equiv \Delta m_{2 1}^{2}
\simeq \lambda^2 v^2 \ee

with \cite{epjc}

\be
\Delta m_{2 1}^{2} \simeq 7 \times 10^{-5} eV^{2}
\ee

A more complicated expression can be found for $1.3 \times 10^{-3}
eV^{2} \leq \Delta_{3 2}^{2}(= m_{3}^{2} - m_{2}^{2}) \leq 3.0
\times 10^{-3} eV^{2}$ \cite{epjc}. Eq.(\ref{matrix}) serves to
predict in some sense the $\lambda's$, $h's$ and $\eta's$, so we
have:

\be
\lambda \simeq 3.4 \times 10^{-14}
\ee

\be
1.8 \times 10^{-2} eV \leq h \leq 2.6 \times 10^{-2} eV
\ee

Now in the mirror sector it is known \cite{palash} that the
hadronic decays of right-handed neutrinos, that is $N \rightarrow
e^{\pm} + hadrons$, are relevant above the pion threshold, which
means $m_{N} \geq 150 MeV$. On the other hand, a typical mirror
scale is $\hat{v} \simeq 10^{3} - 10^{4} GeV$, which gives:

\be
\eta \geq 10^{-4} - 10^{-5}
\ee

if $m_{N}$ is generation independent.

For light neutrinos we have the following masses:

\be
m_{\nu_1} \simeq 8.4 \times10^{-3} eV
\ee

\be
m_{\nu_2} \simeq 1.2 \times 10^{-2} eV
\ee

\be
3.8 \times 10^{-2} eV \leq m_{\nu_3} \leq 5.6 \times 10^{-2}
eV
\ee

and for the mirror neutrinos

\be
m_{N_1} = m_{N_2} \simeq \frac{\eta \hat{v}}{\sqrt{2}}
\ee

\be
m_{N_3} \simeq \sqrt{2} \eta \hat{v}
\ee

Taking $\eta$ of order one we obtain $m_N$ of order
$10^{3}-10^{4}$ $GeV$.

\section{Conclusions}
In this paper we have used a left-right symmetric model with two
scalar doublets in order to generate neutrino masses. We have
written formulae for these masses (standard and mirror) in terms
of free and general parameters of the model $\lambda$, $h$ and
$\eta$, as well of the corresponding vacuum expectation values. In
first instance one could take the matrix $D$ as zero (meaning
absence of mixing among mirror-mirror neutrinos), but due to the
full unitarity of the mixing matrix $U$, this leads us to an also
zero matrix $A$, which is inconsistent since $A$ mixes the
standard neutrinos among themselves. Another possibility is to try
with a realistic matrix $A$ as dictated by experiment, as worked
for instance by King and Mohapatra \cite{king}; however, such
choice is not appropriate because although $A$ itself is unitary,
it breaks as a consequence the complete unitarity of $U$.

Finally, we may diagonalise directly the  matrix $M^{+} M$ with
the help of the mirror scale $\hat{v} \simeq 10^{3} GeV$ (other
scales \cite{coutinho} raises $\hat{v}$ up to $10^{10} GeV$) and
of the experimental predictions for the squared-neutrino mass
differences. In this way we found $\lambda \simeq 10^{-14}$, $\eta
\geq 10^{-4}-10^{-5}$. Needles to say, although we have succeded
on predicting these bounds, more effort and data are required to
get better information and deep understanding on the content of
such parameters, which we hope can help to clarify new physics
beyond the Standard Model.

\section{Acknowledgments}
The authors R. G., A. H. G. and J. M. R. R. wish to thank to
Sistema Nacional de investigadores, SNI, for partial support. J.
M. R. R. would like to thanks Departamento de Matem\'atica
Aplicada de la Universidad Polit\'ecnica de Valencia for the nice
environment during his sabbatical period in this institution.

\end{document}